# A dual origin for water in carbonaceous asteroids revealed by CM chondrites

Laurette Piani[1*†], Hisayoshi Yurimoto[1] and Laurent Remusat[2]

[1]Department of Natural History Sciences, Faculty of Science, Hokkaido University, Japan (*corresponding author: laurette.piani@univ-lorraine.fr)

[2]IMPMC, UMR CNRS 7590 - Sorbonne Universités - UPMC - IRD - MNHN, France

[†]Present address: CPRG, UMR 7358 CNRS, Université de Lorraine, 54500 Vandoeuvre-lès-Nancy, France.

**Carbonaceous asteroids represent the principal source of water in the inner Solar System and might correspond to the main contributors for the delivery of water to Earth. Hydrogen isotopes in water-bearing primitive meteorites, e.g. carbonaceous chondrites, constitute a unique tool for deciphering the sources of water reservoirs at the time of asteroid formation. However, fine-scale isotopic measurements are required to unravel the effects of parent body processes on the pre-accretion isotopic distributions. Here we report in situ micrometer-scale analyses of hydrogen isotopes in six CM-type carbonaceous chondrites revealing a dominant deuterium-poor water component ($\delta D = -350 \pm 40$ ‰) mixed with deuterium-rich organic matter. We suggest that this D-poor water corresponds to a ubiquitous water reservoir in the inner protoplanetary disk. A deuterium-rich water signature has been preserved in the least altered part of the Paris chondrite ($\delta D_{Paris} \geq -69 \pm 163$ ‰) in hydrated phases possibly present in the CM rock before alteration. The presence of the D-enriched water signature in Paris might indicate that transfers of ice from the outer to the inner Solar System have been significant within the first million years of the Solar System history.**

Determining the source(s) of hydrogen, carbon and other volatile elements accreted on Earth is essential for understanding the origins of water and life, and for constraining the chemical and dynamic processes that operated during the evolution of the early Solar System and planet formation (1). Meteorites, micrometeorites and cometary materials can be used in this respect as remnants of the primordial reservoirs of volatile elements present in the protoplanetary disk 4.57 billion years ago. It is however necessary to take into account the effects of alteration processes that occurred within their parent bodies (e.g. hydrothermal alteration and thermal metamorphism), but the modifications induced by these secondary processes on the pre-accretion components remain unclear (2-4).

The CM-type carbonaceous chondrites constitute a well-defined chemical group of meteorites that may be one of the dominant sources for the late veneer of volatile-rich material on the early Earth (5). These chondrites present very variable extents of hydrothermal alteration, and have been described as highly altered to only partially altered (6-8) with petrological subtypes comprise between 2.0 and 2.7/2.9 according to the scale defined by Rubin et al. (6). All CM chondrites contain significant quantities of hydrated minerals and organic matter (9-11), the latter found both as a complex macromolecular carbonaceous component insoluble in water and organic solvents (the so-called insoluble organic matter) and as a mixture of various soluble compounds including molecules of potential biological interest (12). Micrometer-to-nanometer-sized organic matter aggregates are mostly found intertwined at a sub-micrometer scale in the chondrite fine-grained matrix in association with hydrated minerals (13). This small-scale association is challenging for the complete separation and exhaustive analysis of the volatile-bearing components (14-16).

Hydrogen isotopes can be used to track the chemical processes involving water and/or organic matter on parent bodies and to constrain the origin of accreted materials on asteroids and planets (17). By performing whole rock isotope measurements, Alexander et al. (11) showed the presence of positive and distinct correlations between D/H and C/H ratios in both CM- and CR-type carbonaceous chondrites. These correlations indicate that (i) the bulk hydrogen isotopic composition of a given chondrite is the product of mixing a D-poor and C-poor component (presumably the water from which the hydrated minerals formed) with a D-richer organic material in particular proportions, and (ii) that the hydrogen isotope intercepts correspond to the isotopic compositions of the initial water components. The isotopic composition of the water was found to be distinct in each group: $D/H_{initial\ water} = [87 \pm 4] \times 10^{-6}$ or $\delta D_{initial\ water/SMOW} = -444 \pm 23$ ‰ (1σ) for CM chondrites and $D/H_{initial\ water} = [171 \pm 17/10] \times 10^{-6}$ or $\delta D_{initial\ water/SMOW} = 96 \pm 110/-65$ ‰ (1σ) for CR chondrites (where SMOW refers to Standard Mean Ocean Water; *Methods*). These measurements provided important constraints on the nature of the water in the inner protoplanetary disk, which was found to be less D-rich than previously reported. However, bulk measurements such as these do not allow any chemical exchange between water and organics on the asteroid to be decrypted, while measurement of the present (post-alteration) hydrogen isotopic compositions of hydrated minerals and organics within each chondrite remains challenging due to their intimate mixing in the matrix.

We measured the D/H and C/H ratios in the fine-grained matrices of six CM chondrites in situ at a micrometer scale by secondary ion mass spectrometry (SIMS). Depending on the location of the in situ measurement in the matrices, the D/H and C/H ratios may vary as a function of the relative amount of hydrated minerals to organic matter. If, as expected from bulk and insoluble organic matter measurements in CMs (3, 9), the D/H ratios are locally higher in the organics than in the water, then the D/H and C/H ratios should show a positive correlation that reflects the local variations in the proportions of hydrated minerals to organics. These correlations could then be used to estimate the D/H ratio of water for each chondrite as demonstrated in a previous study (16). By comparing the D/H ratios of water in CM chondrites that exhibit different degrees of alteration, we should be able to: (i) examine the extent of water-organic isotope exchanges (e.g. if the D/H ratio of water depends on the relative amount of D-rich organics versus hydrated minerals), (ii) discuss the conditions under which the alteration took place, and (iii) assess the D/H ratios of the primordial volatile reservoirs of the protoplanetary disk.

**Results.** Six CM chondrites covering a wide range of degrees of aqueous alteration were selected from samples that had not experienced significant terrestrial weathering (Table 1 and *Methods*). The selected meteorites cover the complete range of bulk hydrogen contents (reported as equivalent $H_2O$ content) and almost the entire range of bulk isotopic compositions reported for non-anomalous and non-heated CM chondrites (D/H = 120 $\times 10^{-6}$ to 174.5 $\times 10^{-6}$ (11, 22); $\delta^{17}O$ = -2.15 to 4.2‰ and $\delta^{18}O$ = 2.43 to +15‰; highest values measured in the CM chondrite North West Africa 8534 reported in the online <u>Meteoritical Bulletin Database</u>).

The D/H ratios of inorganic hydrogen in individual CM chondrites were derived from in situ measurements by SIMS on matrix areas of 10 x 10 $\mu m^2$ size (*Methods*). The measured D/H ratios are linearly correlated with the measured $^{13}C/H$ ratio, a proxy for the organic matter content of the matrix for each CM chondrite (Figure 1 and Supplementary Figure 2). The zero intercept of the linear fit was used to estimate the D/H ratio of the inorganic H-bearing end-member in each chondrite (*Methods*, Table 2 and Supplementary Table 2). The D/H ratios of the inorganic H-bearing end-member in all CM chondrites except Paris are similar within error, with an average value and 2σ standard deviation of D/H$_{CM}$ = [101 ± 6] $\times 10^{-6}$ (or δD$_{CM/SMOW}$ = -350 ± 40 ‰). The organic end-member is expected to have a C/H ratio two orders of magnitude higher than that measured in the matrix ($^{13}C/H^- \approx 0.1$ for the standard organics; Supplementary Table 1); thus the uncertainty arising from extrapolation to such a high C/H value would be so large that it would be difficult to compare the composition of the organic end-member in Paris with other CM chondrites (Supplementary Figure 3). The inorganic H-bearing end-member in Paris is significantly enriched in deuterium compared to water in the other CM chondrites (Figure 1a and 1b), with the least altered lithology corresponding to D/H$_{Paris}$ = [145 ± 25] $\times 10^{-6}$ or δD$_{Paris/SMOW}$ = -69 ± 163 ‰ after correction for the instrumental fractionation (*Methods*; Table 2). The more altered Paris lithologies are largely D-poor relative to the least-altered lithology. In the D/H ratio vs. C/H ratio plot, these D-poor data show wide scatter without any clear correlation and lie in the space between the trends defined by the D-enriched least-altered lithology and the more altered CM chondrites (Figure 1b).

**Discussion.** The observed positive correlations between the D$^-$/H$^-$ and C$^-$/H$^-$ ratios in each chondrite indicate the presence of two main H-bearing components in the CM matrices: a D-poor/C-poor hydrated phase and a D-rich/C-rich component (Figure 1). FeMg serpentines and cronstedtites, the most abundant hydrated minerals in the CM chondrite matrix (25), probably constitute the major hydrated phases that correspond to the D-poor/C-poor hydrated end-member. In the least altered lithologies of the Paris chondrite, hydrated amorphous silicate materials could also contribute to the hydrogen budget of the C-poor end-member. Hydrated amorphous silicate materials are indeed present with phyllosilicates in the pristine CR3.0 chondrite MET 00426 ($H_2O$ up to 10 wt. %; 26). Hence, the D-rich/C-rich end-member likely corresponds to a mixture of insoluble organic grains and diffuse organic matter, which is finely mixed with hydrated phases in the matrix of CM carbonaceous chondrites (13). At equilibrium, the hydrated silicate isotope composition can be directly used as a proxy of the D/H ratios of water from which the hydrated minerals formed (27; *Methods*).

The hydrogen isotopic composition of CM water (D/H$_{CM}$ = [101 ± 6] ×10$^{-6}$ or δD$_{CM/SMOW}$ = -350 ± 40 ‰ at 2σ, Table 2) is lower than the whole rock and the matrix D/H ratios (D/H ratios of 125 ×10$^{-6}$ to 151 ×10$^{-6}$, 11, 23; Table 1). This confirms the significant contribution of the D-rich organic matter to the total hydrogen budget of the CM chondrites. The hydrogen isotopic composition we measured for CM water is also lower than the compositions deduced by mass balance calculations using the whole rock and insoluble organic matter hydrogen abundances and isotope ratios, e.g., D/H = 126 ×10$^{-6}$ to 142 ×10$^{-6}$ (3, 28). The precision of the mass balance calculations therefore suffers from underestimation of a D-rich component due to the omission of some soluble organic compounds and uncertainties in the amount of insoluble organic matter.

Although the CM chondrites Murchison, Murray, Mighei, Cold Bokkeveld, and Sayama cover a large range of alteration degrees (Table 1), the water from which the hydrated minerals formed retained a unique isotopic composition (Figure 2a). Only the least altered CM chondrite Paris has a significantly higher D/H ratio for its hydrated phases. The absence of any correlation between the water hydrogen isotope composition and the alteration index indicates that no significant isotope exchange or chemical reaction between fluids and other H-bearing phases such as organic matter occurred during alteration. While the oxygen isotope composition of water becomes lighter due to progressive exchange with anhydrous silicates (24) (Table 1), the hydrogen isotopic composition of hydrated silicates remains fixed from the initial stages of the alteration process. In contrast with the water composition, the D/H ratios of the CM matrices and whole rocks show a clear decrease with increasing degree of alteration of the chondrite (Figure 2b). These characteristics and the constant D/H ratio of the water (Figure 2a) argue for dilution of an initially D-rich reservoir in the CM chondrite protolith by a D-poor water component, likely accreted as ice, as the main process responsible for the hydrogen isotope evolution during alteration. The amount of D-poor water would therefore have been higher in the chondrites that show a higher degree of alteration, consistent with the model proposed by Eiler and Kitchen (23) although these authors proposed a subsequent loss of volatile elements during the formation of secondary phases. Indeed, the amplitude of variations for other volatile elements (e.g. N, Ar) appears too large to be simply explained by the addition of a few percent of water (8, 11, 21, 23). The D-poor water component is strongly depleted in deuterium. The co-variations of D/H and C/H ratios in all chondrite matrices (this study) and between the CM whole rocks (11) indicate that the initial D-rich reservoir was dominated by an organic (C-rich) material. Variations in the conditions of the aqueous alteration event (temperature, oxygen fugacity, pH, volatile loss etc.) do not appear to have significantly affected the hydrogen isotopic compositions of the individual components.

The higher D/H ratio for hydrated phases in Paris is surprising given the absence of any isotopic variation of water between Murchison, Murray, Mighei, Cold Bokkeveld and Sayama (Figure 2a). Nevertheless, the hydrogen isotopes in Paris are not completely decoupled from those in other CM chondrites; the more altered areas in Paris display D/H ratios that are scattered between the ratios measured in the least altered lithologies of Paris and those in the more altered CM chondrites (Figure 1). Similarly, C-rich areas ($^{13}C/H^- > 0.015$) are absent in the altered lithologies of Paris, where the C/H ratios are within the range of those measured in other CMs (Figure 1). The comparison of the two types of lithologies in Paris therefore shows a decrease of the D/H and C/H ratios with a local increase in alteration on the parent body. Thus, the D-enrichment in the hydrated phases of the least-altered chondrite Paris seems to have been inherited from the CM protolith present before the aqueous alteration of the CM parent body by a D-poor water.

The amount of water measured in the carbonaceous chondrites whole rock is considered to be a proxy for the alteration degree (11). Even if part of the water may have been lost since the accretion of the parent body, the water content of CMs seems to correlate with both the degree of alteration and the whole rock D/H ratio (Figure 3). The bulk $H_2O$ content is notably lower in Paris than in the other CM chondrites (Table 1). Unlike other CM matrices containing mainly phyllosilicates, the least-altered parts of the Paris matrix are dominated by micron-sized amorphous silicate units with nano-inclusions of Fe-sulfides occasionally edged by a fine fibrous material and separated by porosity (30). The preservation of these characteristics indicates that the matrix has been neither significantly altered nor thermally modified and that no pervasive water has circulated in these least-altered parts (30). It thus appears that the hydrogen isotope signatures measured in the least-altered lithologies of Paris were acquired at the time of or before accretion and have been partly blurred by dilution with D-poor water in the more altered parts. This is consistent with recent oxygen isotope measurements in the Ca-carbonates showing the presence in Paris of a primordial carbonate trend not observed in most other CM chondrites (22). The two other chondrites in which a distinct isotope signature for carbonates was measured are Maribo (31)–a CM2 chondrite having an $^{16}$O-rich bulk oxygen isotope composition close to the one of Paris (32)–and LON 94101 (33)–a CM2 being relatively water-poor comparing to other CMs ($H_2O$ = 8.3 wt.% (11)). This finding was interpreted as the evidence for two water sources having distinct oxygen isotope signatures in Paris (22). In the present study, we confirm the presence of a dual origin of water in the least-altered CM chondrite Paris and show that the two water sources have distinct hydrogen isotope signatures. The oxygen and hydrogen isotope signature of the initial water appears to have been erased in most of the other CM chondrites by the high quantity of D-poor alteration water.

The D-poor alteration water measured in the five altered CM chondrites did not evolve with aqueous alteration on the CM parent body (Figure 2a) and is consistent with that deduced from bulk hydrogen isotope measurements of about 50 CM chondrites (11). This implies that this D-poor composition reflects the main water component accreted on CM chondrite parent body. Interestingly, water with similar D/H ratios has been found in numerous objects of the inner Solar System: in chondrules in the Semarkona ordinary chondrite (OC), the Renazzo CR carbonaceous chondrite, and the Paris CM chondrite ($\delta D_{SMOW} \approx$ -520/-200 ‰; 34-35); in lithic clasts from the Isheyevo carbonaceous chondrites ($\delta D_{SMOW} \approx$ -450/-350 ‰; 36); in Antarctic micrometeorites ($\delta D_{SMOW} \approx$ -370 ‰; 37); in interplanetary dust particles ($\delta D_{SMOW} \approx$ -200 ‰; 38); in phyllosilicates in CR chondrite matrices ($\delta D_{SMOW} \approx$ -220 ‰; 15), and in eucrite and angrite differentiate meteorites ($\delta D_{SMOW} \approx$ -220 ‰; 39-40). These values correspond to the D-poor end-member of arrays of D/H measurements but it is unlikely that it was derived from isotopic fractional of D-rich water as a result of chemical processes; kinetic processes such as distillation and isotopic fractionation at equilibrium with the H-bearing organic phases would enrich the water in heavy isotope. The D-poor water might yet represent one of the main water reservoirs of the inner Solar System and the main reservoir of water in the CM chondrite accretion region.

The D-poor water in CM chondrites could have originated from isotopic re-equilibration in the inner disk between gaseous $H_2O$ and the D-depleted solar $H_2$ (41, 42). This would occur in the disk at temperatures above the water condensation temperature (~160 K). Two factors would control the isotope fractionation between gaseous $H_2O$ and $H_2$: the equilibration temperature (43) and the exchange reaction rate that is very slow below 200 K (44). Given the isotopic composition of the protosolar $H_2$ (45), the fractionation coefficient between gaseous $H_2O$ and $H_2$ (43) and between gaseous and condensed $H_2O$ (46), the D/H ratio we measured for the CM chondrite water could result from water equilibrated with $H_2$ at 350 K and subsequently condensed at 160 K. Another possibility is that the CM water results from the addition of water equilibrated with $H_2$ at temperatures higher than 350 K and subsequently condensed and mixed with D-rich water ice that came from the outer Solar System (47).

The water of hydrated phases in Paris is enriched in deuterium compared to water in other CM chondrites. Its D/H ratio is similar within error to the D/H ratio of water of the CR-type carbonaceous chondrite (11) and could thus originate from the same reservoir. Nonetheless, CR chondrites display variable D/H ratios at a micrometer scale in their coarse-grained hydrated silicates and fine-grained matrix (up to D/H = $400 \times 10^{-6}$), which could result from secondary processes on the CR parent body, including redox reaction with metallic phases (15). In the ordinary chondrite Semarkona, considerable D/H variations (from 150 to 1800 $\times 10^{-6}$) over a scale of a few microns were observed in the hydrated minerals (16). These variations were interpreted as reflecting the onset of localized aqueous alteration occurring at a micron-scale by the melting of isotopically heterogeneous ice inherited from the interstellar medium (14, 16). In Paris, the presence of an isotopically heterogeneous ice would be difficult to reconcile with the systematic decrease in D/H ratios observed in the more altered lithologies. It has been suggested that oxidation of iron by water on the parent body might have enriched the remaining water in D due to $H_2$ escape and Rayleigh-type distillation in slightly altered carbonaceous chondrites (3). To increase the D/H ratio from $101 \times 10^{-6}$ to $145 \times 10^{-6}$ at 100°C, at least half of the water would have to be lost by the least-altered lithology of Paris, a proportion which appears enormous considering the low degree of alteration of these lithologies, the absence of any pervasive water fingerprints (30) and the presence of non-oxidized Fe-Ni metal grains (7). A more encouraging hypothesis would be that D-rich water, predating secondary processes on the CM parent body, was trapped within the CM protolith before accretion and was distinct from the water accreted as ice that led to the CM parent body alteration features. The presence of pre-accretion hydrous minerals (platy crystallized

Fe-cronstedtites) in CM chondrites has indeed been proposed based on their modal mineralogies (25). The hydrous silicates might have formed in the nebula as a by-product of chondrule formation in shock waves within icy regions (48) possibly from a similar reservoir to CR chondrite water. Recent experimental results also demonstrate that amorphous forsterite may have been hydrated into serpentine and brucite in the presence of water vapor within a short time scale in the protoplanetary disk (<10 Myr; 49). The Paris D-enriched signature of hydrated phases is possibly linked to the $^{16}$O-poor signatures observed in some Paris carbonates and could originate from the outer disk ice and preserve inheritance from the molecular cloud (50). Transfers of water from the outer to the inner disk must therefore have been significant before accretion of the CM chondrite parent body within the first million years after CAI formation.

## Main References

**Acknowledgments.** The authors are grateful to the French National Museum of Natural History (Paris) and B. Zanda for providing the pieces of the Paris chondrite, to F. Robert for providing the samples of Murchison, Murray, and Mighei, to the Japanese Museum of Natural History and S. Yoneda for providing the Sayama sample, and to V. Vinogradoff for providing some of the insoluble organic matter isolated from Paris. H. Naraoka from the Planetary Trace Organic Compounds research center is thanked for the measurement of the whole rock $H_2O$ content and D/H ratio of Sayama. F. Baudin from the French Institut des Sciences de la Terre (ISTeP, UPMC-Université Paris 06) is thanked for the measurement of the bulk carbon content of Paris. N. Kawasaki, Y. Marrocchi, B. Marty, N. Sakamoto, I. Sugawara, S. Tachibana, and A. Williams are warmly thanked for fruitful discussions and for providing assistance that allowed this work to be completed. This work was supported by the grant-in-aid for Scientific Research on Innovative Areas "Evolution of molecules in space from interstellar clouds to proto-planetary nebulae" supported by the Ministry of Education, Culture, Sports, Science & Technology, Japan (Grant number 16H00929, L. Piani). This is CRPG contribution #2562.

**Author contributions.** L.P. designed the study, analyzed the samples and wrote the paper. L.R. and H.Y. were involved in the study design and interpretation of the data and also provided input to the manuscript.

**Author information.** Correspondence and requests for materials should be addressed to L. Piani (email: laurette.piani@univ-lorraine.fr).

**Methods**.

**Bulk analysis of Sayama and Paris.** Sayama D/H ratio and water content were determined at the research center for Planetary Trace Organic Compounds (PTOC) at Kyushu University using a chromium filled elemental analyzer coupled to an isotope-ratio mass spectrometer (IRMS) following the method described in (51). The bulk carbon content of the Paris chondrite was measured with a Thermo-Fisher Flash 2000 CHNS-O analyzer at the Institut des Sciences de la Terre (ISTeP, UPMC-Université Paris 06) using three aliquots of ∼ 2 mg of bulk Paris powder, following a method described in (52).

**Sample selection.** Six CM chondrites covering a wide range of degrees of aqueous alteration were selected for this study (Table 1): Sayama (from the Japanese Museum of Natural History, Tokyo), Cold-Bokkeveld, Mighei, Murray and Murchison, and Paris (from the French National Museum of Natural History, Paris). The degree of aqueous alteration of these chondrites as estimated from petrological features (6) increases with their bulk (or whole rock) hydrogen contents (reported as weight percent $H_2O$), their enrichment in heavy oxygen isotopes, and is inversely correlated with their bulk D/H ratios (Table 1). Among the meteorites studied, Sayama is the most altered. Sayama has been described as heavily altered on the basis of petrographic observations (chondrules almost entirely replaced by serpentine and chlorite, presence of abundant pendlandtite, absence of tochilinite, etc.), which is consistent with its heavy oxygen isotope composition and high water content (18, 20; Table 1). Five of the six CM chondrites corresponded to observed falls (Sayama, Cold-Bokkeveld, Mighei, Murray and Murchison). Contrary to the other selected CM chondrites, the Paris chondrite is not an observed fall. Paris has been found in a box bought at an auction in Paris containing belongings of a mining engineer in Africa and SE Asia. Several features of the Paris chondrite indicate that the meteorite was not exposed to significant terrestrial weathering and that it was possibly collected right after its fall: (i) Paris has a very fresh black shiny fusion crust (7), (ii) it contains amino acids and hydrocarbons - molecules sensitive to terrestrial weathering and contamination – showing no signs of contamination, and that are likely better preserved than in other CM falls (53), (iii) no depletion in Na was observed in Paris suggesting that it was collected before being exposed to rain (7, 32).

**Sample preparation**. For in situ isotope measurements, sub-millimeter pieces of each chondrite were handpicked under a stereomicroscope and pressed in pure indium foils. As the Paris chondrite contains local differences in its alteration degree (7), both fresh and more altered areas were obtained from the French National Museum of Natural History and were mounted separately. Fine-grained matrix areas were then localized using a polarized reflected-light microscope and backscattered electron images and chemical maps using dispersive X-ray spectroscopy performed using a JEOL JSM 7000F scanning electron microscope (SEM) at Hokkaido University. After SIMS measurement, the positions of the analyzed areas were also monitored by SEM and points falling outside of the fine-grained matrix areas were rejected.

**Reference materials.** A type III kerogen, the insoluble organic matter of the Antarctic Grosvenor Mountains 95502 (GRO 95502) ordinary chondrite (54), and the insoluble organic matter of the carbonaceous chondrites Murchison, Paris and Orgueil (52) were used as reference materials for organics. A montmorillonite smectite from Camps Bertaux, Morocco (54) and two serpentines (23) were chosen as references for hydrated minerals. Supplementary Table 1 summarizes the reference compositions and the measured isotopic and elemental compositions of the standard samples. Following our previous works (16, 54), laboratory-prepared mixtures of the D-rich insoluble organic matter from GRO 95502 and the D-poorer montmorillonite were also analyzed during a separate session to test our protocol. All samples and standards were pressed in indium and gold-coated before analyses.

**Analytical conditions.** The Secondary Ion Mass spectrometry (SIMS) measurements were performed with the IMS-1280HR installed at Hokkaido University during two different sessions of measurements, in December 2015 and October 2016. A 10 keV $Cs^+$ primary beam was used for the measurements. Prior to analyses, 8 minutes of a high-current (1.5 nA) pre-sputtering was applied in order to clean the surface and reach the sputtering steady-state. The samples were then sputtered with an 80 to 250 pA beam shaped by an aperture in the primary column allowing a large homogeneous ellipsoidal shape with a major axis of about 70 μm to be obtained. A normal incident electron gun was used to improve charge compensation. The analyzed area was restricted to 10 x 10 $μm^2$ in the center of the shaped beam by using a small field aperture and a high magnification mode. The $H^-$, $D^-$, $^{13}C^-$ and $^{29}Si^-$ ions were collected successively by changing the magnetic field and counted with the monocollection electron multiplier. The primary current was adjusted to obtain about 1.5 x $10^5$ c/s on $H^-$ with primary currents of 250 pA for the chondrite matrices and hydrous mineral references and 80 pA for the standards of organic matter. The mass resolution power was set to M/ΔM = 3300 to avoid interferences on $^{13}C^-$ and $^{29}Si^-$. For each analysis, 50 cycles were collected with 1 s of counting time per cycles for $H^-$, $^{13}C^-$ and $^{29}Si^-$ and 10 s for $D^-$ for a total duration of about 30 minutes. Because carbon isotope variations are limited among the Solar System materials (variation of 10% in the worst case between insoluble organic matter and calcites; (55), lower than the reproducibility in the measurement of the C/H ratio on standards; Supplementary

Table 1) the [13]C/H ratio can be used as a proxy for the C/H ratio. The lower abundance of [13]C comparing to [12]C allows us to measure carbon with an electron multiplier in the whole range of C concentration from the C-poorest to the C-richest matrix areas. Nanoscale carbonates, known to be present in fine-grained matrices might also contribute to the C[-] signal measured by SIMS. Nevertheless, as the carbon content and ionization efficiency of carbonates are low compared to organic carbon (56) and as no deviation from the C/H vs. D/H correlation was observed in the most carbonate-rich areas, their contribution is considered insignificant. No clear relation involving the amount of [29]Si[-] with [13]C[-], H[-] or D[-]/H[-] contents was observed in the matrices. The statistical error on the D/H ratio for the sample having the lowest D/H ratio is 30 ‰ ($2\sigma$) and the reproducibility on the insoluble organic matter of GRO 95502 is of 80 ‰ ($2\sigma$ standard deviation). The D/H ratios estimated for water in chondrites were corrected from the instrumental fractionation using a calibration line determined on standard materials (57) (calibration parameters in Supplementary Table 1). $\delta$D values are calculated as follow: $\delta$D = (D/H$_{sample}$/D/H$_{SMOW}$ - 1) × 1000, with D/H$_{SMOW}$ = 155.76 ×$10^{-6}$ and SMOW referring to as Standard Mean Ocean Water. The uncertainty on the calibration in the D-poor domain, which is not covered by the reference materials, is the main source of uncertainty on the final D/H ratio of chondritic water (about 110 to 185‰ at $2\sigma$). Six to eighty-six areas of fine-grained matrix were measured in each CM chondrite (Supplementary Table 2). Mechanical mixtures of D-rich insoluble organic matter (from GRO 95502 OC chondrite) and the D-poor hydrous silicates (montmorillonite) proportions mimicking the ones in the matrix were also analyzed. We verified that the mixtures and pure components follow a single line in a D[-]/H[-] vs. [13]C[-]/H[-] plot (Supplementary Figure 1) that shows the progressive increase of hydrogen isotope ratio with the amount of organic matter of the analyzed area. A similar trend for organic-richer mixtures of the same components was also reported in a previous work (16).

**Estimation of the water D/H ratio.** The zero intercept of the linear fit (i.e. the D[-]/H[-] ratio at C[-]/H[-] = 0) was used to estimate the D/H ratio of water in each chondrite (Supplementary Table 2). Pearson correlation coefficients range from 0.77 to 0.94, with the most altered chondrites, Sayama and Cold Bokkeveld, having the highest correlation coefficients (Supplementary Table 2). Because the main H-bearing phases in the CM matrices correspond to hydrated silicates (Fe-Mg serpentine and cronstedtite; 13, 25), and as the equilibrium isotope fractionation factor $\alpha$ between serpentine (or other hydrated silicates) and water is lower than the precision of the D/H ratio (1000 × ln $\alpha_{serpentine-water} \approx$ -43 (27) at a alteration temperature of 100°C (58)), we consider the zero intercept to be a direct proxy for the D/H ratio of the water from which the minerals formed. In the Paris least-altered lithologies, where the abundance of phyllosilicates is lower than in the altered parts of the matrix (30) hydrated amorphous silicate materials (26) could also contribute to the H signal of the inorganic end-member. Another possible H-rich/C-poor phases that has been reported in carbonaceous chondrites is ammonia ($NH_3$) but its concentration is several orders of magnitude lower than the water (or H) bulk content (e.g. $NH_3 = 19$ µMol/g in Murchison; 59). Preliminary isotope images obtained with the NanoSIMS 50 at MNHN Paris do not indicate that any N-enrichment in the C-poor domains (Supplementary Figure 4). The H-rich C-poor areas in Paris rather correspond to oxygen-rich domains consistent with water or OH contained in phyllosilicates and/or in the abundant amorphous silicates (30). For Paris, as even the least-altered lithology shows the fingerprints of parent body aqueous alteration (8, 30), and as the alteration water has a lower D/H ratio, the estimated D/H$_{Paris}$ value (Table 2) can be considered a minimum D/H ratio for the Paris pre-accretion water signature.

**Data availability.** The data that support the plots within this paper and other findings of this study are available from the corresponding author upon reasonable request.

**Additional references**

**Figures & Tables.**

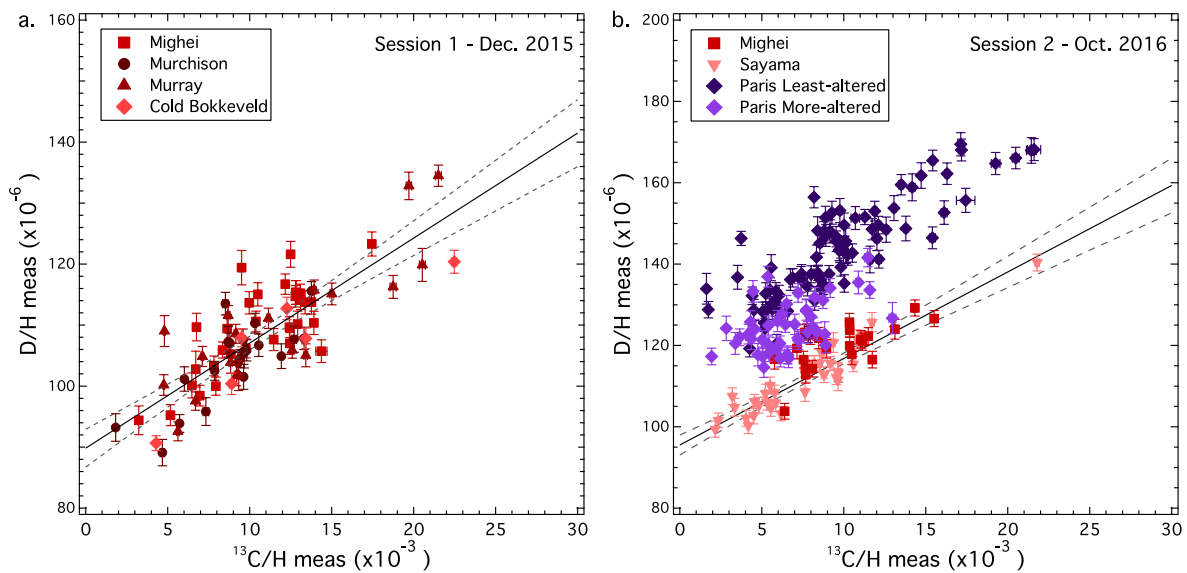

**Figure 1. Measured D/H vs. C/H ratios in the matrices of CM chondrites.** (a) Session 1 data from Murchison, Murray, Mighei, Cold Bokkeveld; linear fit (black line) and 95% confidence interval bands (dashed lines) calculated on all CM data. (b) Session 2 data from with Sayama, Mighei, and Paris least-altered and more-altered lithologies; linear fit (black line) and 95% confidence bands (dashed lines) calculated on Sayama data only. Error bars represent the internal error at $2\sigma$.

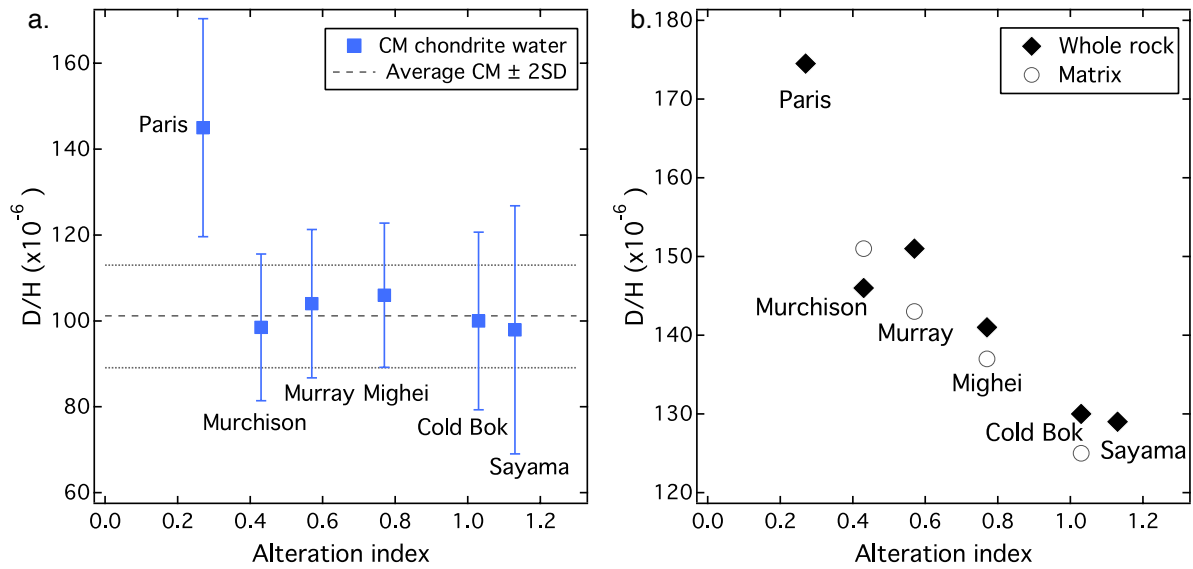

**Figure 2. D/H ratios of water and of whole rock for the measured chondrites as a function of the alteration index** (6, 8). (a) D/H ratios of the in situ water (this study). Error bars (2σ) include the internal error and the error on the instrumental mass fractionation correction calculated from reference materials (see Methods). CM average and standard deviation interval (SD = standard deviation) values were calculated on all chondrites except Paris. (b) D/H ratios measured in whole rocks and in isolated matrix pieces (11, 22, 23 and this study). The alteration index of Sayama has not been determined. As Sayama is considered to be more altered than Cold Bokkeveld according to Chokai et al. (29), we arbitrarily assigned it an alteration index above that of Cold Bokkeveld. Cold Bok = Cold Bokkeveld.

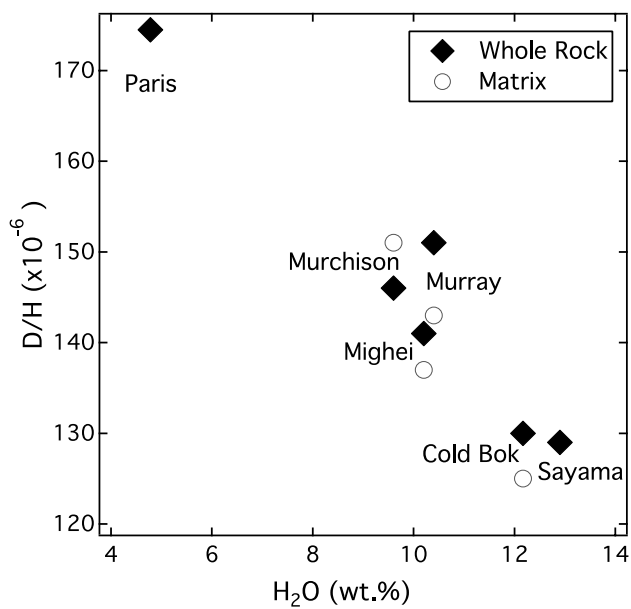

**Figure 3. Whole rock and matrix D/H ratios of the studied CM chondrites as a function of $H_2O$ in weight %.** Data from (11, 22, 23) and this study (Sayama whole rock data). Cold Bok = Cold Bokkeveld.

**Table 1.** Selected chondrites, their petrological type, alteration degree and elemental and isotopic compositions.

| Meteorite | Petrologic subtype | Description | Alteration Index | C (wt.%) | $H_2O$ (wt.%) | C/H$_{WR}$ (at.%) | D/H$_{WR}$ x10$^{-6}$ | D/H$_{Matrix}$ x10$^{-6}$ | δ$^{17}$O$_{WR}$ (‰) | δ$^{18}$O$_{WR}$ (‰) |
|---|---|---|---|---|---|---|---|---|---|---|
| Sayama | CM2 | highly altered[a,e] | >1.03? | 1.99[a] | 12.9[*] | 0.12[a] | 129[*] | - | 3.2/4.2[a] | 9.9/12.0[a] |
| Cold-Bokkeveld | CM2.2[b] | altered[b] | 1.03[g] | 2.45[h] | 12.2[h] | 0.15[h] | 130[h] | 125[j] | 2.76[k] | 10.01[k] |
| Mighei | CM>2.4[b] | partially altered[b] | 0.77[g] | 2.48[h] | 10.2[h] | 0.18[h] | 141[h] | 137[j] | 1.44[k] | 7.57[k] |
| Murray | CM2.4/2.5[b] | partially altered[b] | 0.57[g] | 2.33[h] | 10.4[h] | 0.17[h] | 151[h] | 143[j] | 0.85[k] | 7.53[k] |
| Murchison | CM2.5[b] | partially altered[b] | 0.43[g] | 2.08[h] | 9.6[h] | 0.16[h] | 146[h] | 151[j] | 1.20[k] | 7.30[k] |
| Paris | CM2.7[c,d] | least altered[f,c,d] | 0.27[c] | 1.64[*] | 4.8[i] | 0.26[*] | 174.5[i] | - | -2.15 +0.75[f] | 2.43 +6.80[f] |

WR = whole rock; [*]this study, details in *Methods*; [a]Yoneda et al. 2001 (18); [b]Rubin et al. 2007 (6); [c]Marrocchi et al. 2014 (8); [d]Rubin et al. 2015 (19); [e]Takaoka et al. 2001 (20); [f]Hewins et al. 2014 (7); [g]Browning et al. 1996 (21) [h]Alexander et al. 2012 (11); [i]Vacher et al. 2016 and Erratum (22); [j]Eiler and Kitchen 2004 (23); [k]Clayton and Mayeda 1999 (24)

**Table 2.** D/H ratios of water in the different CM chondrites.

| | Meteorite | D/H$_{water}$ ± 95%CI (×10$^{-6}$) | δD$_{water}$ ± 95%CI (‰) |
|---|---|---|---|
| Session 1 | Murchison | 99 ± 17 | -368 ± 110 |
| | Murray | 104 ± 17 | -335 ± 111 |
| | Mighei | 106 ± 17 | -321 ± 108 |
| | Cold-Bokkeveld | 100 ± 21 | -357 ± 133 |
| Session 2 | Paris all points | 131 ± 24 | -156 ± 154 |
| | Paris least alt. | 145 ± 25 | -69 ± 163 |
| | Paris more alt. | 137 ± 26 | -122 ± 169 |
| | Sayama | 98 ± 29 | -371 ± 185 |
| D/H or δD$_{CM}$ ± 2SD (×10$^{-6}$) | | 101 ± 6 | -350 ± 40 |

D/H$_{water}$ = D/H ratio of chondrite waters after correction of the instrumental mass fractionation; D/H$_{CM}$ = average of the CM water D/H ratio calculated from Murchison, Murray, Mighei, Cold-Bokkeveld and Sayama. CI = confidence interval.